\documentclass[twoside]{article}
\usepackage{graphicx,amssymb,mathrsfs,amsmath}
\input vatola.sty \input cyracc.def
\input psfig.sty

\newcommand{\sanhao}{\fontsize{19.08pt}{3\baselineskip}\selectfont}
\newcommand{\xiaosihao}{\fontsize{12pt}{\baselineskip}\selectfont}
\newcommand{\wuhao}{\fontsize{10.5pt}{\baselineskip}\selectfont}
\newcommand{\xiaowuhao}{\fontsize{9.5pt}{\baselineskip}\selectfont}
\newcommand{\dawuhao}{\fontsize{10pt}{0.8\baselineskip}\selectfont}
\newcommand{\liuhao}{\fontsize{7.875pt}{\baselineskip}\selectfont}

\newcommand{\qihao}{\fontsize{7pt}{\baselineskip}\selectfont}
\newcommand{\bahao}{\fontsize{8pt}{\baselineskip}\selectfont}
\TagsOnRight

\renewcommand{\baselinestretch}{1.06} 
\catcode`@=11 \long\def\@makefntext#1{\noindent #1}
\newskip\tabcentering \tabcentering=1000pt plus 1000pt minus 1000pt
\def\REF#1{\par\hangindent\parindent\indent\llap{#1\enspace}}%\ignorespaces}
\def\MCH#1#2{\setbox0=\hbox{\raise#1\hbox{#2}}\smash{\box0}}% move char

\def\@evenfoot{}\def\@oddfoot{}
\def\@evenfoot{\vbox{\hbox to \textwidth{\bahao\sf\hbox to
0.01cm{\textbf{\thepage}\hfill} \hfill{\emph{????? et al. Sci China
Ser G-Phys Mech Astron } {$|$ Dec. 2008 $|$ vol. 51 $|$ no. 12 $|$
\textbf{1100-1120}} }\hfill}}}
\def\@oddfoot{\vbox{\hbox to \textwidth{\bahao\sf\hbox to
0.01cm{} \hfill{ \emph{\hspace{8mm}????? et al. Sci China Ser G-Phys
Mech Astron } {$|$ Dec. 2008 $|$ vol. 51 $|$ no. 12 $|$
\textbf{1100-1120}} }\hfill\hfill\textbf{\thepage}}}}
\def\sec#1{\vspace{6mm}\noindent{{\xiaosihao\sf\textbf{#1}}}\vspace{2mm}}% ¶¨ÒåÒ»¼¶Õ½Ú
  \def\tlj{\end{document}}  \newsymbol\wjzhml 203F

\def\wj{\end{document}}
\def\no{\noindent}

\begin{document}
\abovedisplayskip=5pt plus 1pt minus 2pt %¹«Ê½ÒÔÉϾàÀë
\belowdisplayskip=5pt plus 1pt minus 2pt %¹«Ê½ÒÔϾàÀë
%-------------------  First Head  -----------------------------------------
%-------------------  First Head  -----------------------------------------
\textwidth=145truemm \textheight=212truemm
\renewcommand{\baselinestretch}{0.9}\baselineskip 9pt
{\psfig{figure=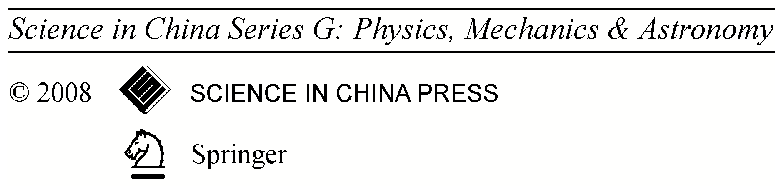}\hfill
\begin{picture}(43,0)
\rightline{\sf\put(-50,25){{\vbox{\hbox {\hspace{5.85mm}\dawuhao
www.scichina.com} \hbox{\hspace{6.5mm}\dawuhao phys.scichina.com}
\hbox{\,\dawuhao www.springerlink.com}}}}}
\end{picture}
%-------------------  First Head  -----------------------------------------
%-------------------  First Head  -----------------------------------------

\vspace{11.6true mm}
\renewcommand{\baselinestretch}{1.6}\baselineskip 19.08pt
\noindent{\sanhao{\sf\textbf{Models and Observations of Sunspot\\Penumbrae}}}
\vspace{0.5 true cm}
\renewcommand{\sfdefault}{phv}

\noindent{\sf BORRERO, Juan Manuel
\footnotetext{ \baselineskip 6pt \qihao
\vspace{-2.2mm}\\
Received ; accepted \\
doi \\
$^\dag$Corresponding author (email: borrero@mps.mpg.de)\\

\centerline{\bahao\sf \emph{Sci China Ser G-Phys Mech Astron} $|$
Dec. 2008 $|$ vol. 51 $|$ no. 12 $|$ \textbf{1100-1120}}
}}

\vspace{0.2 true cm}\noindent
\parbox{13.3cm}
{\noindent\renewcommand{\baselinestretch}{1.3}\baselineskip 12pt
{\liuhao\sf Max Planck Institut f\"ur Sonnensystemforschung, Max Planck Strasse 2, D 37191, Katlenburg-Lindau, Germany\vspace{2mm}}}

\noindent{\xiaowuhao\sf\textbf{\hspace{-1mm}
\parbox{13.3cm}
{\noindent
\renewcommand{\baselinestretch}{1.3}\baselineskip 13pt
The mysteries of sunspot penumbrae have been under an intense scrutiny for the past 10 years. 
During this time, some models have been proposed and refuted, while the surviving ones 
had to be modified, adapted and evolved to explain the ever-increasing array of observational 
constraints. In this contribution I will review two
of the present models, emphasizing their contributions to this field, but also
pinpointing some of their inadequacies to explain a number of recent observations at very high
spatial resolution. To help explaining these new observations I propose some modifications to each of 
those models. These modifications bring those two seemingly opposite models closer together into a
general picture that agrees well with recent 3D magneto-hydrodynamic simulations.}}}

\vspace{5.5mm}\no{\footnotesize \sf% ¹Ø¼ü´Ê
Sunspots, Magnetic fields, Spectropolarimetry, Magnetohydrodynamics}\vspace{6mm}
\baselineskip 15pt

\renewcommand{\baselinestretch}{1.08}
\parindent=10.8pt  %\parskip=2mm
\rm\wuhao\vspace{-4mm}

%%%----------------------------------------------------%%%
\sec{1\quad Embedded flux-tubes and Field-free gap models.}

In its most basic manifestation the structure of the penumbral magnetic field can be
described as being {\it uncombed} (i.e. composed of two distinct interlaced components). 
The first component, {\it spines}, is characterized by a strong ($B \sim 1700$ Gauss) and 
inclined ($\gamma \sim 45^{\circ}$)\footnote{$^{1}$~Here $\gamma$ refers to the inclination of
the magnetic field with respect to the vertical direction on the solar surface.} 
magnetic field. Because of the similarities it shares with the umbral magnetic field,
this component is often thought to be an extension of it. The second component, 
{\it instraspines}, appears interlocked in between spines and is characterized by a weaker 
and more horizontal ($B \sim 1200$ Gauss, $\gamma  \sim 90^{\circ}$) magnetic field. This 
spine-intraspine structuring is already seen at 1 arcsec resolution ([1]). 
The most successful models to explain the uncombed penumbral structure are the embedded flux-tube
and the field-free gap models.

The idea of penumbral flux tubes is an early concept gathered when first stratospheric balloons
 took continuum images of sunspots at 1 arcsec in the late 50's and early 60's\footnote{$^{2}$~There 
is a very brief reference to {\it tubes} in Danielson (1960,[2]).} revealing in detail the 
filamentary structure of the penumbra. However, it was not until Solanki \& Montavon (1993,[3]) 
that penumbral flux-tubes were first invoked
to explain the uncombed structure of the penumbral magnetic field. In this model, penumbral intraspines 
are assumed to be composed by at least one horizontal magnetic flux-tube. The Evershed 
flow is channeled along these radially aligned flux-tubes, which are embedded in a surrounding atmosphere 
with a less inclined and generally stronger magnetic field (i.e. spines).

The Field-free gap model (hereafter referred to as {\it gappy penumbra}) was possibly first proposed by
Choudhuri (1986,[4]) to explain the connection between umbral dots and penumbral bright grains. It was 
later employed to explain the uncombed nature of the penumbral magnetic field ([5]).
In this picture, field-free plasma rising from the beneath the sunspot would pierce into the sunspot magnetic field
from below, creating a region right above the field-free gap where the magnetic field is horizontal and weaker
(intraspine) than in the gap's surroundings, where the field is stronger an less inclined (spines).

\sec{2\quad The Evershed flow}

Within the frame of the flux-tube models the Evershed flow is usually explained in terms of a siphon flow 
channeled along a thin flux-tube ([6]). This concept was developed in detail for steady ([7]) 
and dynamic ([8],[9]) flux tubes. Simulations based on siphon flows help to explain 
supercritical and supersonic Evershed velocities ([10]), formation of shock fronts ([11]), 
proper motions of penumbral grains ([12],[13]), and moving magnetic features as a continuation of the 
Evershed effect in the sunspot canopy ([14],[15],[16],[17],[52],[56]).

The gappy penumbral model, as initially proposed ([4],[18]), does not clearly identify the origin of 
Evershed effect. Within the field-free gap there are convective flow motions (upflows at the gaps' 
center and downflows on the edges), but these are unable to explain the radially outwards Evershed flows observed in the 
deep photosphere. It has been proposed ([18]) that the hot plasma inside the gap could
 heat the atmosphere above it, where the horizontal field lies, producing a local version of the mechanism 
introduced by Schlichenmaier et al. (1998,[8]) to drive the Evershed flow.

\sec{3\quad Spectropolarimetric observations}

Both the flux-tube and gappy-penumbra models feature one of the key ingredients to explain the polarization
profiles observed in the penumbra at $\simeq 1$ arcsec resolution: the uncombed structure of the magnetic field 
(see Sect.~1). The other requirement is the presence of a strong ($\gtrsim 5$ km s$^{-1}$) radial outflow in the weaker 
and horizontal component of the magnetic field. This is readily included in the flux-tube model (see Sect.~2), and
consequently this model is able to explain the anomalous polarization signatures observed in sunspot penumbrae
(e.g. multi-lobed and highly asymmetric Stokes $V$ profiles). This includes the Stokes profiles in the visible
 Fe I 630 nm ([19]), near-infrared Fe I 1.56 $\mu$m ([11],[20],[21],[22]), Ti 2.2 $\mu$m spectral lines ([23])
and even simultaneous observations in different wavelength ranges ([53],[54],[55]). It is also possible to explain the 
azimuthal variation of the net circular polarization ([24],[25],[26],[27]), as well at is center-to-limb variation 
([3],[26],[28]) in different spectral lines.

So far, the gappy penumbral model has not be used to explain the polarization signals in the
sunspot penumbra. Without further modifications, such a attempt will face severe difficulties 
since this model does not include a radial outflow in the horizontal magnetic field above the 
field-free gap. Having a horizontal flow along the field-free gap (instead or in addition to the convective 
flow) will have little effect  because this flow must be present within a magnetized atmosphere.

\sec{4\quad Dark-cored penumbral filaments and penumbral heating}

The high average brightness of the penumbra (about 70 \% of the quiet Sun) imposes strong constraints as to which
 mechanism is responsible for its heating ([29]). Within the horizontal flux-tube model, the
energy source invoked is the Evershed flow itself, which appears in the inner penumbra (where
the flux-tubes are slightly tilted upwards) as a hot upflow that develops inside the flux tube 
and quickly becomes horizontal. This hot upflow heats the filament as it moves radially outwards. Numerical
estimates have shown that the radiatively cooling time is very short, producing bright filaments
that are much shorter than observed ([30]). Schlichenmaier \& Solanki (2003,[31]) postulate that 
long and bright filaments can be explained if a new hot upflow
appears right after the previous one cools down and sinks as a downflow. However, the smooth
variations of the inclination observed along filaments at very high spatial resolution seem to
rule out this possibility ([32],[33]). This problem has been revisited recently ([34])
generalizing the calculations done by Schlichenmaier et al. (1999,[30]) to three dimensions,
thick flux-tubes\footnote{$^{3}$~ With a radius larger than the pressure scale-height in the Photosphere 
(about 100 km)}, and embedded in a magnetized atmosphere with a more realistic temperature stratification. 
It has been possible to reproduce considerably longer bright filaments, so that the Evershed flow remains
as a possible heating source. In this work they have also explained, in terms
of opacity effects, the dark lanes observed by Scharmer et al. (2002,[35]) at the core of penumbral 
filaments. The required densities and temperatures are in agreement with 
those necessary to keep a thick flux-tube in magnetohydrostatic equilibrium with the surrounding
magnetic atmosphere ([36],[37]).

Contrary to the problems faced by the horizontal flux-tube model, the gappy penumbral model
allows for a very efficient heating mechanism. Similar to granulation, convective flows inside 
the field-free gaps could provide the energy to maintain the penumbral brightness. The continuous upflow
 along the full length of the filament carries much more energy than the localized (mainly in the inner
 penumbra) upflows at the flux-tubes' inner footpoints. Furthermore, the hot upflow at the center of the 
gap would create a density enhancement at its top. This locally rises the continuum level
where the plasma is cooler, thus producing a central dark lane. Although no calculations have been carried 
out to confirm either effect, this appears to be a plausible scenario.

\sec{5\quad Magnetic field inclination}

Results from the study of the observed polarization signals in the penumbra have revealed 
that the magnetic field vector of the component carrying the Evershed flow (intraspines) 
is tilted slightly upwards in the inner penumbra: $\gamma \sim 70^{\circ}$, whereas it 
is directed downwards in the outer penumbra. $\gamma \sim 110^{\circ}$. This is seen both 
at low ([11],[19],[21]) and high spatial resolution ([38]). 

This would represent no major obstacle if it was not for the fact that these
regions are often as large as $\sim 2-4$ Mm radially. A flux-tube in the inner
penumbra pointing $\sim 70^{\circ}$ with respect to the vertical would rise more
than 700 km along that distance, quickly escaping from the layers where spectral lines
are formed.

The gappy penumbral model does not suffer from this shortcoming in the inner penumbra, since 
the vertical component of the magnetic field does not totally vanish on top of the gap.
This yields magnetic fields slightly tilted upwards there, while the gap beneath can 
remain horizontal and thus remaining close to the line-forming layers. However,
this peculiarity of the gappy penumbral model poses problems in the outer penumbra,
where inclinations of $\sim 110^{\circ}$ with respect to the vertical are also observed
over radially extended regions. This would imply a field-free gap that sinks more than
700 km in just 2-3 Mm, thus escaping from the line-forming layers. The situation is aggravated 
because this model predicts inclinations close to $90^{\circ}$ only 
on a very small region right above the gap. In fact, near $\tau_5=0.1-0.01$ the inclinations 
values are closer to $70-80^{\circ}$, forcing the gap to sink even further to explain the 
observed inclinations.

Note that the magnetic field inside flux-tubes can be twisted and thus it can posses 
inclinations larger than $90^{\circ}$ while the tube's axis remains horizontal ([37]). 
Therefore flux-tube models do not suffer from this problem in the outer penumbra.

\sec{6\quad An unifying picture from 3D MHD simulations}

Very recently, the first 3D MHD simulations of a sunspot have been presented ([39],[40]).
So far these simulations have been restricted to grey radiative transfer and a moderate
 grid separation (20-30 km). Despite these shortcomings they have been able to reproduce a 
number of features that resemble the penumbral structure as seen from continuum images: 2-3 
Mm long penumbral filaments featuring dark-cores and lifetimes of about 1 hour. 

Figure 1 shows the properties of the magnetic field and velocity vectors in a vertical 
slice across one of the simulated filaments by Rempel et al.([40]). In this figure $V_x$ 
will be referred to as radial velocity or Evershed flow. The subsurface 
structure of these filaments reveals plumes of weak and horizontal field below the visible surface 
of the penumbra\footnote{$^{4}$Inside the plume, the magnetic field is not totally horizontal 
(inclined about $\gamma \simeq 60-70^{\circ}$). Together with the fact that the simulated 
penumbra is a factor of 2-3 smaller than typically observed, this makes these simulations mostly representative
of the inner penumbra.}. The plumes carry an upflow at its center that turns over near the $\tau_5=1$ level
and feeds downflows along the sides of the plume. These results share common points with both the 
flux-tube model and the gappy penumbral model. The magnetic field inside the plume is highly inclined 
due to the expulsion of the vertical component of the magnetic field due to convective motions. However these
motions have little effect on the horizontal component of the field.

\begin{enumerate}
\item The plumes' typical vertical extension ($\sim 1$ Mm) is much larger than its horizontal extension 
($\sim 300$ km). This does not support the concept of a \emph{round} flux-tube, although vertically 
elongated flux-tubes are still possible. This possibility is interesting because makes the flux-tube
more stable against the action of the external field ([36]). However, this same effect rules out
the possibility of detecting the lower boundary of the flux-tube ([19]).
\item Plumes do not reach the bottom of the simulation box ($\sim 6$ Mm), nor they originate 
in the convection zone, but rather they form within the surrounding field. This supports the concept of 
embedded flux-tube as opposed to a field-free gap piercing from beneath the sunspot.
\item The plumes contain a horizontal field of about 750-900 Gauss. This is inconsistent with the 
concept of a \emph{field-free} gap ([41]). However, numerical simulations with higher resolution
tend to yield smaller magnetic fields.
\item Plumes sustain a convective flow pattern (Fig.~1; lower panels) in terms of a hot upflow across its 
center and cool downflows along the plumes' edges. These convective motions in the deep layers are
in agreement with the predictions of the gappy penumbral model ([41]). This is also supported by a number 
of recent observations ([43],[44],[45]), although other observations at disk center with better spatial and
spectral resultion do not find them ([57]).
\item Because the field is not totally horizontal inside the plume, the downflows has a component 
towards the umbra that presents itself as an inverse Evershed flow. At the top of the plume the 
flow turns horizontal ($V_x$ peaks) deflected by the highly inclined magnetic field there and producing 
a flow pattern near $\tau_5$ that resembles the Evershed flow. No indications of siphon flows are found.
\item In these simulations the mechanism responsible for the energy transport and heating of the penumbra is mainly
performed by the convective flow in the $YZ$-plane. This is in better agreement with the gappy penumbral model. However, 
it does not support the simulations based on thin flux-tubes, where the energy  transport is produced by the hot upflowing gas
 in the inner penumbra that is transported along the $X$-axis (tube's axis). 
\end{enumerate}

\begin{figure*}
\begin{center}
\begin{tabular}{cc}
\includegraphics[width=8cm]{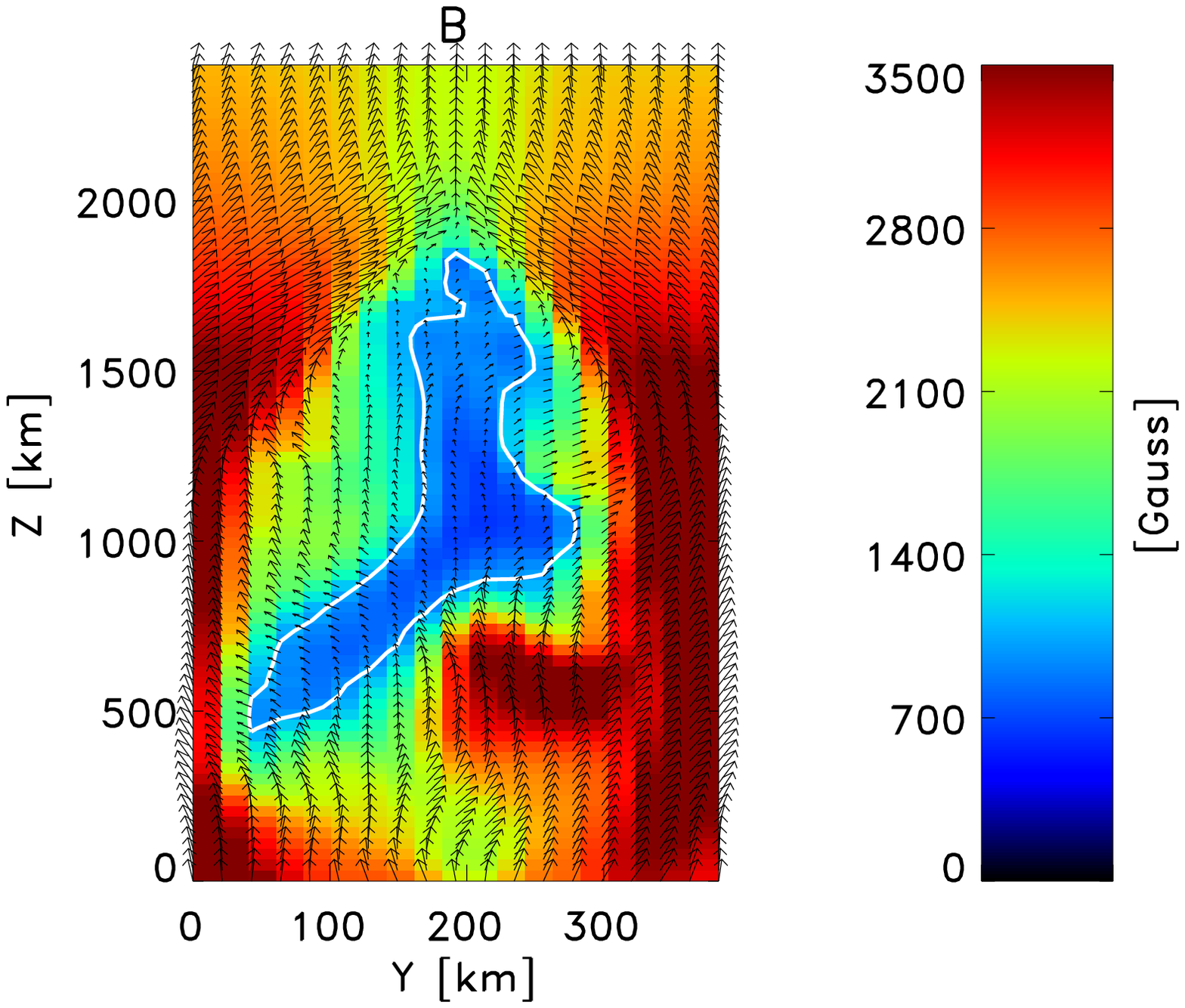} &
\includegraphics[width=8cm]{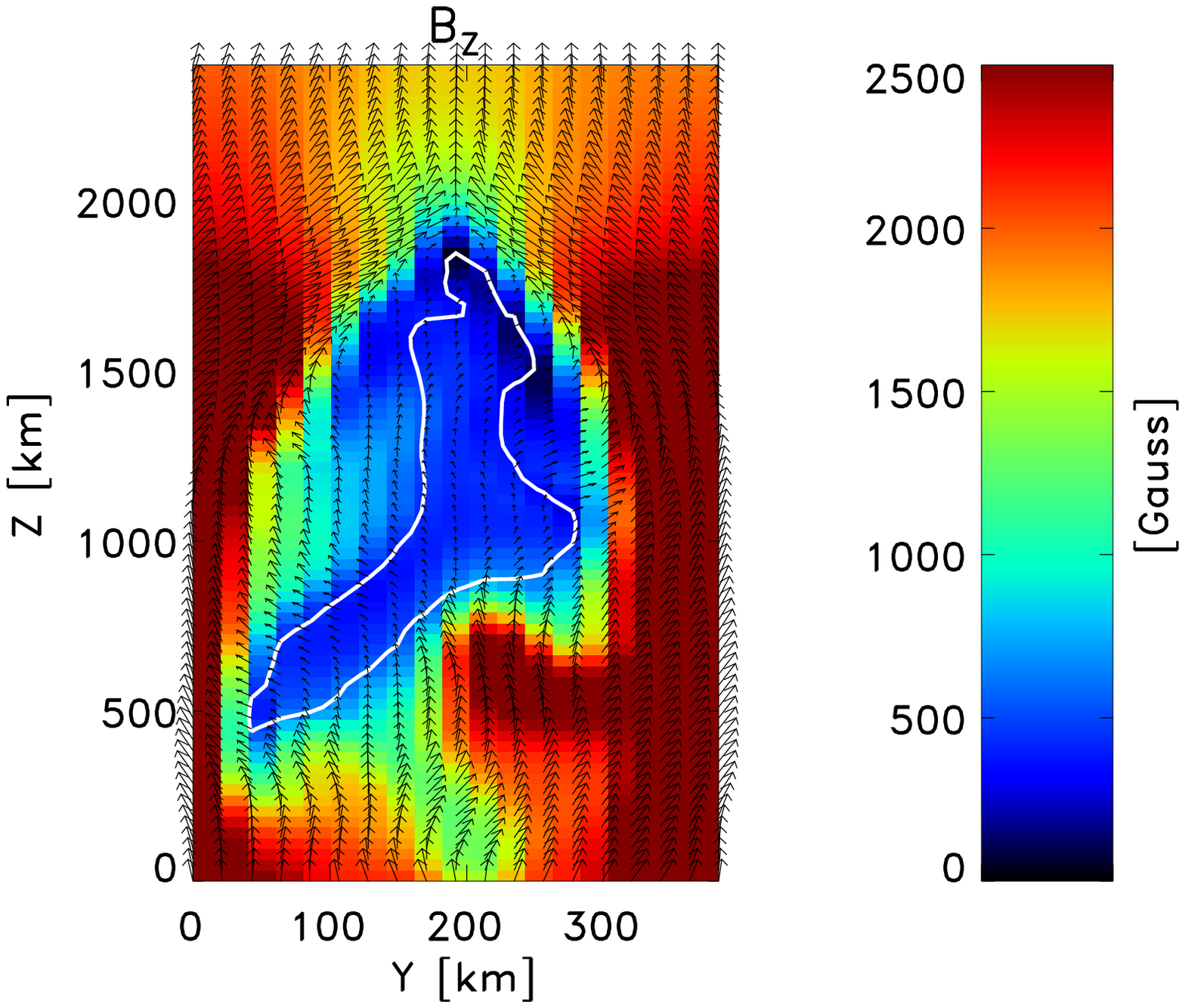} \\
\\
\includegraphics[width=8cm]{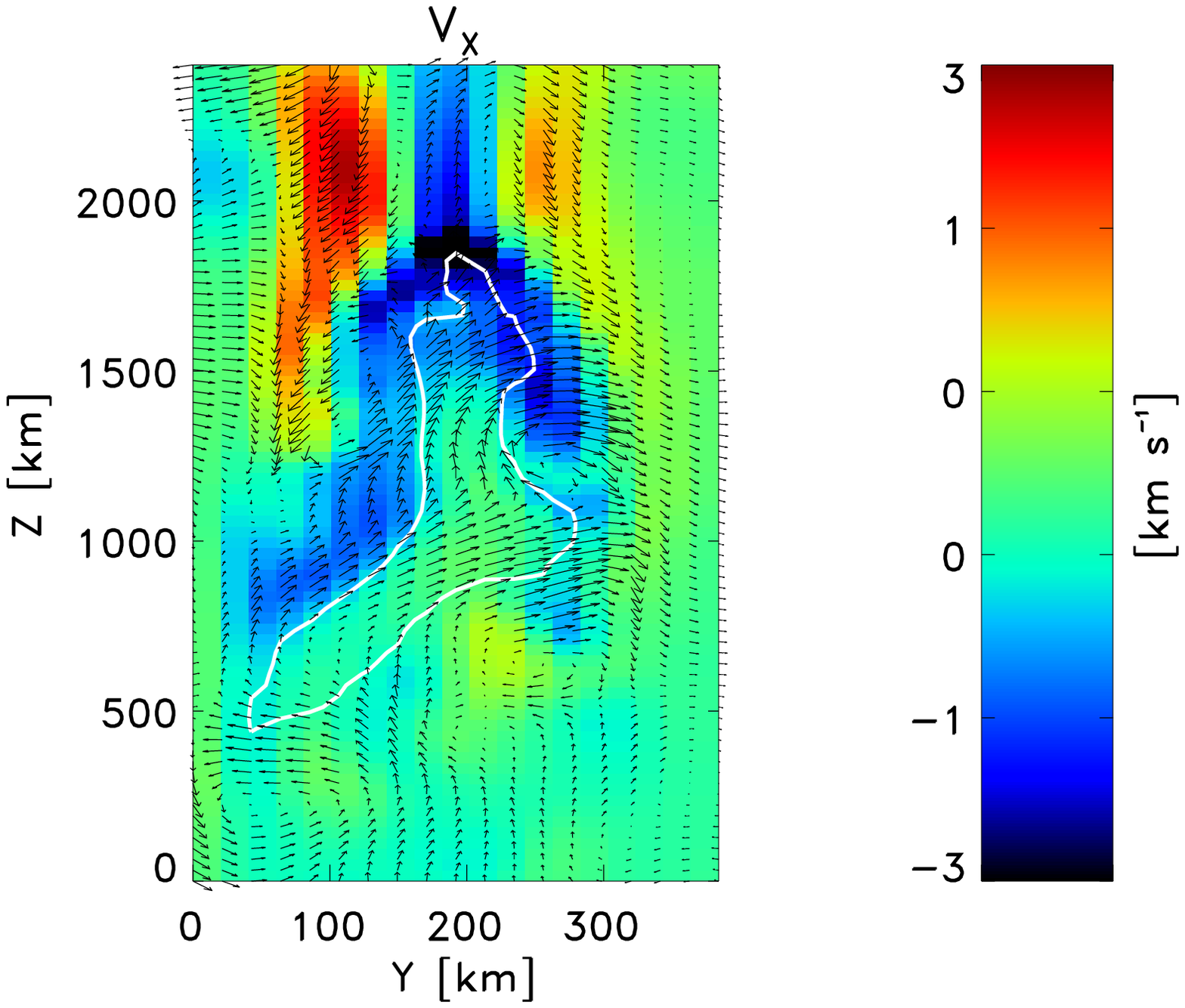} &
\includegraphics[width=8cm]{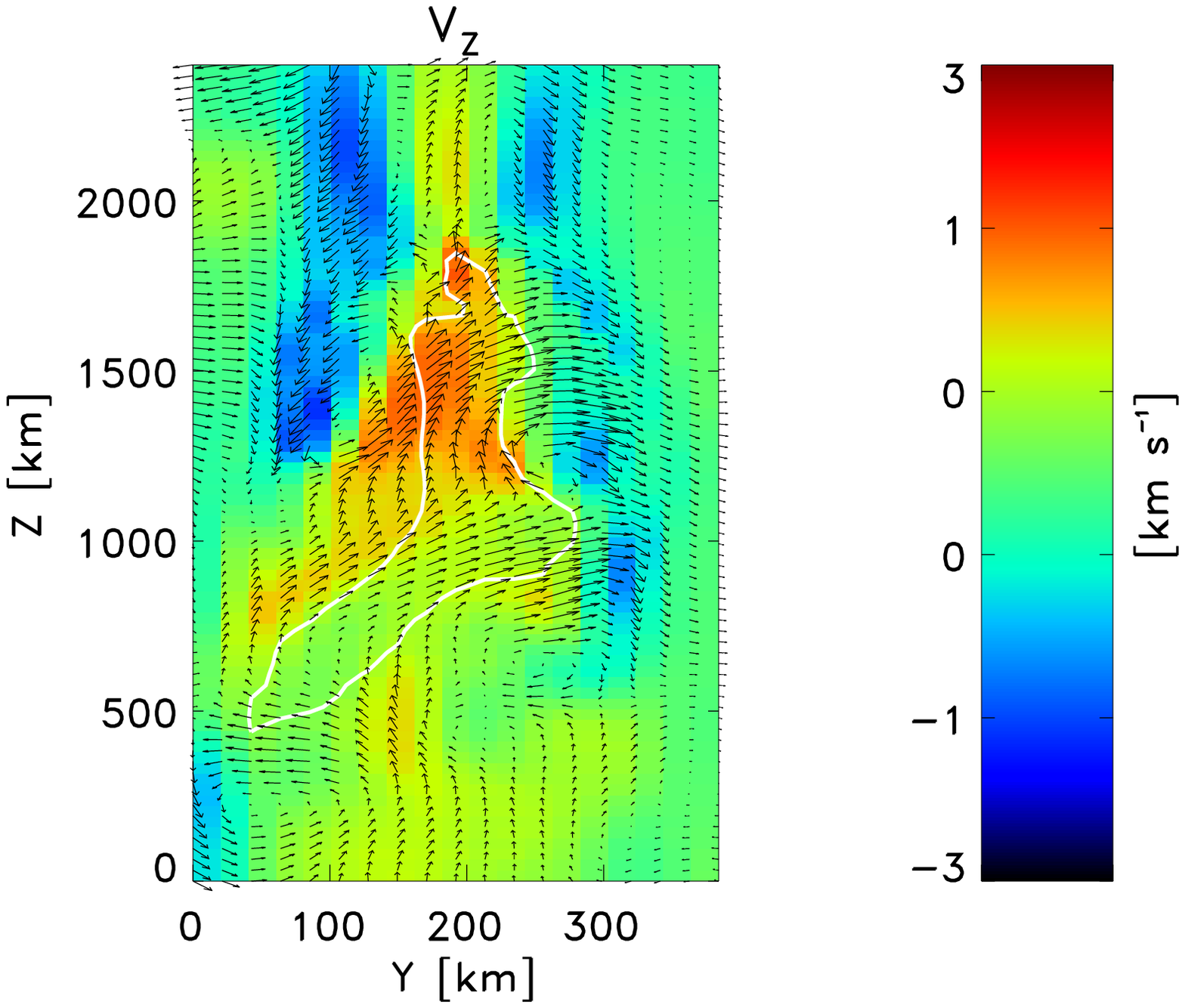} \\
\end{tabular}
\caption{Vertical slice across a penumbral filament from a snapshot in Rempel et al's simulations. The $X$-axis 
corresponds to the radial direction along the penumbra. {\it Upper-left}: total magnetic field. {\it 
Upper-right}: vertical component of the magnetic field $B_z$. The white contour encloses the region where the 
total field strength is smaller than 1000 Gauss. The mean magnetic field inside this area is about 850 Gauss. 
The arrows show the magnetic field vector in the $YZ$-plane. {\it Lower-left}: radial component of the
velocity $V_x$ (radial direction in the penumbra). {\it Lower-right}: vertical component of the
velocity $V_z$. The arrow field shows the velocity field in the $YZ$-plane.}
\end{center}
\end{figure*}

Keeping these simulations in mind, we could bring together the gappy and embedded-flux tube model if
on the one hand we consider a flux-tube (where the magnetic field is horizontal) that is highly
squeezed vertically and that harbors a convective flow pattern (besides the horizontal 
Evershed flow along its axis). The energy transported by the convective motions would 
explain the heating of the penumbra (Sect.~4). On the other hand, we could take a field-free 
gap (harboring a convective flow pattern) and fill it with a horizontal magnetic field of
 about 1000 Gauss and a horizontal velocity of at least 5 kms$^{-1}$. This would allow the gappy 
penumbral model to explain the polarization profiles observed in the penumbra (Sect.~3).

\sec{7\quad Open questions}

Despite the success of these 3D MHD simulations there are a number of 
issues that remain unclear. Many of them will certainly be solved as simulations
become more realistic and observations of better quality become available. However, some 
others cast doubt as to whether these simulations represent the real magnetoconvective 
process occurring in the penumbra. To address these concerns and others, it is compulsory 
to carry our forward modeling of Stokes profiles using non-grey simulations.

A first concern has to do with the convective velocity pattern along the edges of the plumes,
which yields velocities that are directed inwards instead of outwards. This inflow is yet to
be discovered. Note that the inflow predicted by these simulations is not exactly the same
as the flow pattern reported by Zahkarov et al. (2008,[43]). Both show downflows at the
edges of the filaments ($V_z<0$), but in Zahkarov et al. the Evershed outflow ($V_x>0$) is overposed 
(but canceled in their data by observing perpendicularly to the line of symmetry of the sunspot), 
whereas in the simulations is directed inwards ($V_x<0$).  Perhaps the inflow remains hidden because
better spatial resolution is needed to detect it, or because it occurs below the elevated $\tau_5=1$ 
thus remaining invisible.

The same argument concerning the location of the $\tau_5=1$ level is often used to support the claim
that plumes tend to be more field-free in simulations with higher resolution (Sect.6, item 3).
Then, why do we not observe regions free of magnetic field ([41]) ? Again, the solution invoked 
is the opacity effect at the top of the plume, that rises the $\tau_5=1$ level to a region which is not totally 
field-free ([48]).

However, this explanation remains controversial since in these simulations $\tau_5=1$ is
formed above umbral dots, whereas in penumbral filaments it is formed inside it (see also Fig.~3 
in [5]). This is due to the larger difference in the magnetic field strength between plumes and their surroundings 
in the umbra, which yields a larger gas pressure and thereby rising the $\tau_5=1$ level in umbral dots as 
compared to penumbral filaments. Therefore, if convective motions and field-free regions are difficult to
observe in penumbral filaments, it should be even more difficult in the case of umbral dots. However,
both convective motions ([46],[47]), and almost field-free regions (down to $\sim 400$ Gauss,[47]) are indeed
seen in umbral dots. The only possible explanation we are left with, is that the penumbral plumes \emph{are not} field-free,
whereas umbral plumes \emph{are}. The reason why these simulations yield smaller fields inside penumbral plumes
for higher resolution runs might be because these simulations are not representative of the mid- or outer- penumbra 
(where the magnetic field is almost horizontal; see footnote 4), and thus convective motions
are still relatively effective in getting rid of the magnetic flux inside the plume, through the expulsion
of the still strong vertical component of the field.

Another very problematic issue, coming back to the observed velocities, is the ubiquitousness of 
the Evershed outflow. No region in the penumbra seems to be free of it, as it appears inside 
intraspines but also weakly in spines (Fig.~14 in [49], Fig.~2-3 in [50]). None of the mentioned 
models, including numerical simulations, can explain these observations.

More concerns with the velocity field appear as we realize that the convective flow
has a similar magnitude compared with the Evershed flow ($\sim 2-3$ kms$^{-1}$). This 
vigorous convection is clearly needed to bring enough energy from deep below into the 
photosphere and heat the penumbra. On the one hand, the convective velocities are too 
strong to have remained unseen for so long. A possible explanation is that near 
$\tau_5=1$ the vertical velocities are smaller: ($\lesssim 1$ kms$^{-1}$). On the other hand, 
the Evershed flow in these simulations is clearly a factor 2-3 weaker than observed. Indeed, 
it is commonly found that the Evershed flow carries supercritical velocities (at least 30 \% of the penumbra,[21]). 
If indeed the vertical (convective) velocities are much smaller than the radial velocities, 
then the energy input associated to the Evershed flow can have a non-negligible contribution to the heating 
of the penumbra.

Another important point of discrepancy is related to the inclination of the magnetic field inside
the plumes. Fig.~1 (top-right panel) shows that $B_z$ inside the plume does not totally vanish
(these simulations are representative of the inner penumbra, see footnote 4). It is not clear what process 
will switch $B_z<0$ inside the plumes to explain the large regions exhibiting flux return in the outer penumbra.
Perhaps the solution to this riddle will be some sort of magnetic flux pumping occurring inside the plume
at the outer penumbra due to the interaction with the surrounding granulation ([51]).

%\newpage
\normalsize \vskip0.16in\parskip=0mm \baselineskip 15pt
\renewcommand{\baselinestretch}{1.12}
\footnotesize
\parindent=6mm

\bahao\REF{1\ } Lites, B. W.; Elmore, D. F.; Seagraves, P.; Skumanich, A. P.: {\it Stokes Profile Analysis and 
Vector Magnetic Fields. VI. Fine Scale Structure of a Sunspot}. 1993, ApJ, 418, 928
\REF{2\ } Danielson, Robert E.: {\it The Structure of Sunspot Penumbras}. 1960, AJ, 65, 343
\REF{3\ } Solanki, S. K.; Montavon, C. A. P.: {\it Uncombed fields as the source of the broad-band
 circular polarization of sunspots }. 1993, A\&A, 275, 283
\REF{4\ } Choudhuri, A.R.: {\it The dynamics of magnetically trapped fluids. I - Implications for umbral dots 
and penumbral grains}. 1986, ApJ, 302, 809
\REF{5\ } Spruit, H. C.; Scharmer, G. B.: {\it Fine structure, magnetic field and heating of 
sunspot penumbrae}. 2006, 447, 343
\REF{6\ } Meyer, F.; Schmidt, H. U.: {\it Magnetisch ausgerichtete Strömungen zwischen 
Sonnenflecken.}. 1968, ZaMM, 48, 218
\REF{7\ } Montesinos, B.; Thomas, J.H.: {\it The Evershed effect in sunspots as a siphon 
flow along a magnetic flux tube}. 1997, Nature, 390, 485
\REF{8\ } Schlichenmaier, R.; Jahn, K.; Schmidt, H. U.: {\it Magnetic flux tubes evolving in sunspots. 
A model for the penumbral fine structure and the Evershed flow}. 1998, A\&A, 337, 897
\REF{9\ } Schlichenmaier, R.: {\it Penumbral fine structure: Theoretical understanding}. 2002,
AN, 323, 303
\REF{10\ } del Toro Iniesta, J.C.; Bellot Rubio, L.R.; Collados, M.: {\it Cold, Supersonic Evershed 
Downflows in a Sunspot}. 2001, ApJ, 549, 139
\REF{11\ } Borrero, J. M.; Lagg, A.; Solanki, S. K.; Collados, M: {\it On the fine structure 
of sunspot penumbrae. II. The nature of the Evershed flow}. 2005, A\&A, 436, 333
\REF{12\ } Sobotka, M.; Brandt, P. N.; Simon, G. W.: {\it Fine structure in sunspots. III. Penumbral grains}.
1999, A\&A, 348, 621
\REF{13\ } Sobotka, M.; Sütterlin, P.: {\it Fine structure in sunspots. IV. Penumbral grains in speckle 
reconstructed images}. 2001, A\&A, 380, 714
\REF{14\ } Kubo, M.; Shimizu, T.; Tsuneta, S.: {\it Vector Magnetic Fields of Moving Magnetic Features 
and Flux Removal from a Sunspot}. 2007, ApJ, 659, 812
\REF{15\ } Zhang, J.; Solanki, S. K.; Wang, J.: {\it On the nature of moving magnetic feature 
pairs around sunspots}. 2003, A\&A, 399, 755
\REF{16\ } Zhang, J.; Solanki, S. K.; Woch, J.; Wang, J.: {\it The velocity structure of moving magnetic 
feature pairs around sunspots: support for the U-loop model}. 2007, A\&A, 471, 1035
\REF{17\ } Ryutova, M.; Hagenaar, H.: {\it Magnetic Solitons: Unified Mechanism for Moving Magnetic 
Features}. 2007, SoPh, 246, 281
\REF{18\ } Scharmer, G. B.; Spruit, H. C.: {\it Magnetostatic penumbra models with field-free gaps}.
2006, A\&A, 460, 605
\REF{19\ } Borrero, J. M.; Solanki, S. K.; Lagg, A.; Socas-Navarro, H.; Lites, B.:
 {\it On the fine structure of sunspot penumbrae. III. The vertical extension of penumbral filaments}.
 2006, A\&A, 450, 383
\REF{20\ } Bellot Rubio, L. R.; Collados, M.; Ruiz Cobo, B.; Rodríguez Hidalgo, I. 2002,
NCimC, 25, 543
\REF{21\ } Bellot Rubio, L. R.; Balthasar, H.; Collados, M.: {\it Two magnetic components in sunspot 
penumbrae}. 2004. A\&A, 427, 319
\REF{22\ } Borrero, J. M.; Solanki, S. K.; Bellot Rubio, L. R.; Lagg, A.; Mathew, S. K.:
{\it On the fine structure of sunspot penumbrae. I. A quantitative comparison of two semiempirical 
models with implications for the Evershed effect}. 2004, A\&A, 422, 1093
\REF{23\ } R\"uedi, I.; Solanki, S. K.; Keller, C. U.; Frutiger, C.: {\it Infrared lines as probes of solar 
magnetic features. XIV. TI i and the cool components of sunspots}. 1998, A\&A, 338, 1089
\REF{24\ } Schlichenmaier, R.; M\"uller, D. A. N.; Steiner, O.; Stix, M.: {\it Net circular polarization 
of sunspot penumbrae. Symmetry breaking through anomalous dispersion}. 2002, A\&A, 381, L77
\REF{25\ } M\"uller, D. A. N.; Schlichenmaier, R.; Steiner, O.; Stix, M.: {\it Spectral signature of 
magnetic flux tubes in sunspot penumbrae}. 2002, A\&A, 393, 305
\REF{26\ } Borrero, J. M.; Bellot Rubio, L. R.; Müller, D. A. N.: {\it Flux Tubes as the Origin 
of Net Circular Polarization in Sunspot Penumbrae}. 2007, ApJ, 666, L133
\REF{27\ } Tritschler, A.; M\"uller, D. A. N.; Schlichenmaier, R.; Hagenaar, H. J.:
{\it Fine Structure of the Net Circular Polarization in a Sunspot Penumbra}. 2007, A\&A, 671, 85
\REF{28\ } Mart{\'\i}nez Pillet, V.: {\it Spectral signature of uncombed penumbral magnetic fields}. 
2000, A\&A, 361, 734
\REF{29\ } Solanki, S.K.: {\it Sunspots: An overview}. 2003, A\&ARv, 11, 153
\REF{30\ } Schlichenmaier, R.; Bruls, J. H. M. J.; Sch\"ussler, M.: {\it Radiative cooling of a hot flux 
tube in the solar photosphere}. 1999, A\&A, 349, 961
\REF{31\ } Schlichenmaier, R.; Solanki, S. K.: {\it On the heat transport in a sunspot penumbra}. 
2003, A\&A, 411, 257
\REF{32\ } Langhans, K.; Scharmer, G. B.; Kiselman, D.; Löfdahl, M. G.; Berger, T. E.:
{\it Inclination of magnetic fields and flows in sunspot penumbrae }. 2005, A\&A 436, 1087
\REF{33\ } Langhans, K.; Scharmer, G. B.; Kiselman, D.; Löfdahl, M. G.:
{\it Observations of dark-cored filaments in sunspot penumbrae}. 2007, A\&A, 464, 763
\REF{34\ } Ruiz Cobo, B.; Bellot Rubio, L. R.: {\it Heat transfer in sunspot penumbrae. Origin of 
dark-cored penumbral filaments}. 2008, A\&A, 488, 749
\REF{35\ } Scharmer, G.B.; Gudiksen, B.V.; Kiselman, D.; L\"ofdahl, M.G.; Rouppe van der
 Voort, L.H.M.: {\it Dark cores in sunspot penumbral filaments}. 2002, Nature, 420, 151
\REF{36\ } Borrero, J. M.; Rempel, M.; Solanki, S. K.: {\it The uncombed penumbra}. 2006, ASPC, 358, 19
\REF{37\ } Borrero, J. M.: {\it The structure of sunspot penumbrae. IV. MHS equilibrium for penumbral 
flux tubes and the origin of dark core penumbral filaments and penumbral grains}. 2007, A\&A, 471, 967
\REF{38\ } Bellot Rubio, L. R.; Tsuneta, S.; Ichimoto, K.; Katsukawa, Y.; Lites, B. W.; Nagata, S.; 
Shimizu, T.; Shine, R. A.; Suematsu, Y.; Tarbell, T. D.; Title, A.M.; del Toro Iniesta, J.C.: {\it Vector 
Spectropolarimetry of Dark-cored Penumbral Filaments with Hinode}. 2007, ApJ, 668, L61
\REF{39\ } Heinemann, T.; Nordlund, \AA.; Scharmer, G. B.; Spruit, H. C.: {\it MHD Simulations of Penumbra 
Fine Structure}. 2007, ApJ, 669, 1390
\REF{40\ } Rempel, M.; Schuessler, M.; Knoelker, M.: {\it Radiative MHD simulation of sunspot structure}.
2008, ApJ, submitted
\REF{41\ } Borrero, J.M.; Solanki, S.K.: {\it Are there field-free gaps near $\tau=1$ in sunspot penumbrae ?}.
2008, ApJ, in press
\REF{42\ } Scharmer, G. B.; Nordlund, \AA.; Heinemann, T.: {\it Convection and the Origin of Evershed 
Flows in Sunspot Penumbrae}. 2008, ApJ, 677, L149
\REF{43\ } Zakharov, V.; Hirzberger, J.; Riethm\"uller, T. L.; Solanki, S. K.; Kobel, P.:
{\it Evidence of convective rolls in a sunspot penumbra}. 2008, A\&A, 488, L17
\REF{44\ } Ichimoto, K.; Suematsu, Y.; Tsuneta, S.; Katsukawa, Y.; Shimizu, T.; Shine, R. A.; Tarbell, T. D.; 
Title, A. M.; Lites, B. W.; Kubo, M.; Nagata, S.: {\it Twisting Motions of Sunspot Penumbral Filaments}. 2007,
Science, 318, 1597
\REF{45\ } Rimmele, T.: {\it On the Relation between Umbral Dots, Dark-cored Filaments, and Light Bridges}.
2008, ApJ, 672, 684
\REF{46\ } Bharti, Lokesh; Jain, Rajmal; Jaaffrey, S. N. A.: {\it Evidence for Magnetoconvection in 
Sunspot Umbral Dots}. 2007, 665, L79
\REF{47\ } Riethmüller, T. L.; Solanki, S. K.; Lagg, A. {\it Stratification of Sunspot Umbral Dots from 
Inversion of Stokes Profiles Recorded by Hinode}. 2008, ApJ, 678, L157
\REF{48\ } Sch\"ussler, M.; V\"gler, A. {\it Magnetoconvection in a Sunspot Umbra}. 2006, ApJ, 641, L73
\REF{49\ } Bellot Rubio, L. R.; Schlichenmaier, R.; Tritschler, A.: {\it Two-dimensional spectroscopy 
of a sunspot. III. Thermal and kinematic structure of the penumbra at 0.5 arcsec resolution}. 2006, A\&A, 453, 1117
\REF{50\ } Borrero, J. M.; Lites, B. W.; Solanki, S. K.: {\it Evidence of magnetic field wrapping around 
penumbral filaments}. 2008, A\&A, 481, L13
\REF{51\ } Thomas, John H.; Weiss, Nigel O.; Tobias, Steven M.; Brummell, Nicholas H.: {\it
Downward pumping of magnetic flux as the cause of filamentary structures in sunspot penumbrae}. 2002, 
Nature, 420, 390
\REF{52\ } Sainz Dalda, A.; Martínez Pillet, V.: {\it Moving Magnetic Features as Prolongation of Penumbral Filaments}.
2005, ApJ, 632, 1176
\REF{53\ } Cabrera Solana, D.; Bellot Rubio, L. R.; Beck, C.; Del Toro Iniesta, J. C. {\it Temporal evolution of the 
Evershed flow in sunspots. I. Observational characterization of Evershed clouds}. 2007, A\&A, 475, 1067
\REF{54\ } Cabrera Solana, D.; Bellot Rubio, L. R.; Borrero, J. M.; Del Toro Iniesta, J. C. {\it Temporal evolution 
of the Evershed flow in sunspots. II. Physical properties and nature of Evershed clouds}. 2008, A\&A, 477, 273
\REF{55\ } Beck, C. {\it A 3D sunspot model derived from an inversion of spectropolarimetric observations and its 
implications for the penumbral heating}. 2008, A\&A, 480, 825
\REF{56\ } Cabrera Solana, D.; Bellot Rubio, L. R.; Beck, C.; del Toro Iniesta, J. C. {\it Evershed Clouds as Precursors 
of Moving Magnetic Features around Sunspots}. ApJ, 2006, 649, L41
\REF{57\ } Bellot Rubio, L. R.; Langhans, K.; Schlichenmaier, R. {\it Multi-line spectroscopy of dark-cored penumbral filaments}.
2005, A\&A, 443, L7

\tlj